\documentclass[a4paper,11pt]{article}
\usepackage{pos}

\title{An LNGS Mobile Neutron Detector (ALMOND): Mapping Ambient Neutron Background of Gran Sasso National Laboratory}

\author*[a,b]{Melih Solmaz}
\author[c]{Klaus Eitel}
\author[d,e]{Alfredo Davide Ferella}
\author[c]{Felix Kratzmeier}
\author[c,d,e]{Francesco Pompa}
\author[c]{Kathrin Valerius}

\affiliation[a]{Kirchhoff Institute for Physics, Heidelberg University, 69120 Heidelberg, Germany}
\affiliation[b]{Institute of Experimental Particle Physics, Karlsruhe Institute of Technology, 76021 Karlsruhe, Germany}
\affiliation[c]{Institute for Astroparticle Physics, Karlsruhe Institute of Technology, 76021 Karlsruhe, Germany}
\affiliation[d]{Department of Physics and Chemistry, University of L’Aquila, 67100 L’Aquila, Italy}
\affiliation[e]{INFN, Laboratori Nazionali del Gran Sasso, 67100 Assergi, Italy}

\emailAdd{melih.solmaz@kip.uni-heidelberg.de}

\abstract{In deep underground laboratories, environmental neutrons,
which are produced at the cavern walls, introduce a source of background
to rare event searches. The flux and spectrum of the ambient neutrons
vary considerably with time and location. Precise knowledge of
this background is necessary to devise shielding and
veto mechanisms, thereby improving the sensitivity of
the neutron-susceptible underground experiments.
ALMOND, currently in operation, is a low-flux mobile neutron spectrometer
developed for the LNGS underground laboratory to measure the ambient neutron
background of the entire facility. In this paper, an overview of the
design, construction and calibration of ALMOND is given. Furthermore, the result of the first underground neutron measurement is shown along with
an outlook for future measurements and analyses.
}

\FullConference{19th International Conference on Topics in Astroparticle and Underground Physics (TAUP2025)\\
24–30 Aug 2025\\
Xichang, China\\}

\begin{document}
\maketitle

\section{ALMOND project}

Ambient neutrons constitute a source of background for rare event searches
carried out at the LNGS underground laboratory. The majority of them
are produced due to the intrinsic radioactivity in the walls
of the lab cavern~\cite{Wulandari:2003cr}. The uranium-thorium content
of the walls and the water level in the surrounding contribute to
variations in this background. This implies that the ambient neutron
background depends both on the location and time of the measurement.
Previously, various ambient neutron surveys were carried out at LNGS.
However, a direct comparison between them is rather difficult, since
the measurements took place at different locations. In addition,
surveys employed distinct detector technologies characterized
by unique systematic uncertainties and specific target energies.
This adds further complexity to the comparison~\cite{Best:2015yma}.
To address this issue and develop a holistic view of this background,
\underline{A}n \underline{L}NGS \underline{Mo}bile
\underline{N}eutron \underline{D}etector (ALMOND) project was initiated.

ALMOND is essentially made from a stack of plastic scintillator (PS) bars
wrapped with Gd foils~\cite{Solmaz:2023hzr}.
Fast neutrons moderate in PS blocks and cause
proton recoil scintillation. Upon thermalization, they are captured by
the Gd foils. When capture $\gamma$-rays induce an energy deposit (3 MeV
or above) greater than that of the ambient gamma field, neutron
detection is identified with appropriate efficiency. Then,
the prior proton recoils, which are correlated with the neutron
capture, give a measure of incoming neutron energy. This technique is
known as capture-gated neutron spectroscopy~\cite{KAMYKOWSKI1992559}.

\begin{figure}[htbp]
\centering
\includegraphics[width=.48\textwidth]{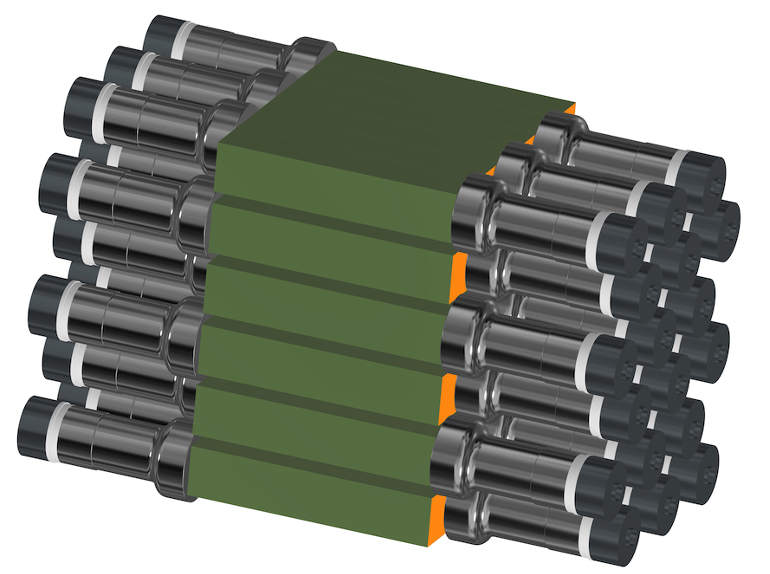}
\includegraphics[width=.45\textwidth]{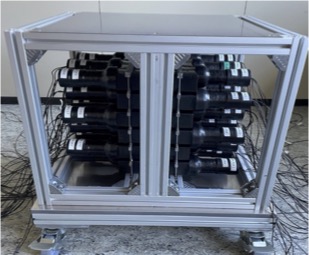}
\caption{(Left) CAD drawing of ALMOND. (Right) Constructed assembly of ALMOND.
The bottom lead plate above the wheels is seen in the picture.}
\label{fig:ALMOND}
\end{figure}

ALMOND consists of 36 individual detector modules that are placed
in a 6x6 arrangement, as illustrated in figure~\ref{fig:ALMOND} (left).
Each segment has a 5 cm x 5 cm x 25 cm EJ-200 PS bar and a 3-inch
9302B type PMT coupled to the PS with optical glue. To improve light collection, the PS is covered with reflector layers. The PS is wrapped
with 100 $\mu$m thick Gd foils in all lateral sides to enhance neutron
sensitivity. To eliminate light leakage, each module is covered with
aluminum foils and then black tape. Figure~\ref{fig:ALMOND} (right) shows
that the entire detector is held by a wheeled support structure, providing
ALMOND with mobility. The support frame incorporates 16 mm thick lead sheets
in all directions to reduce the ambient gamma background.

\section{Calibrations}

\begin{figure}[htbp]
\centering
\includegraphics[width=.48\textwidth]{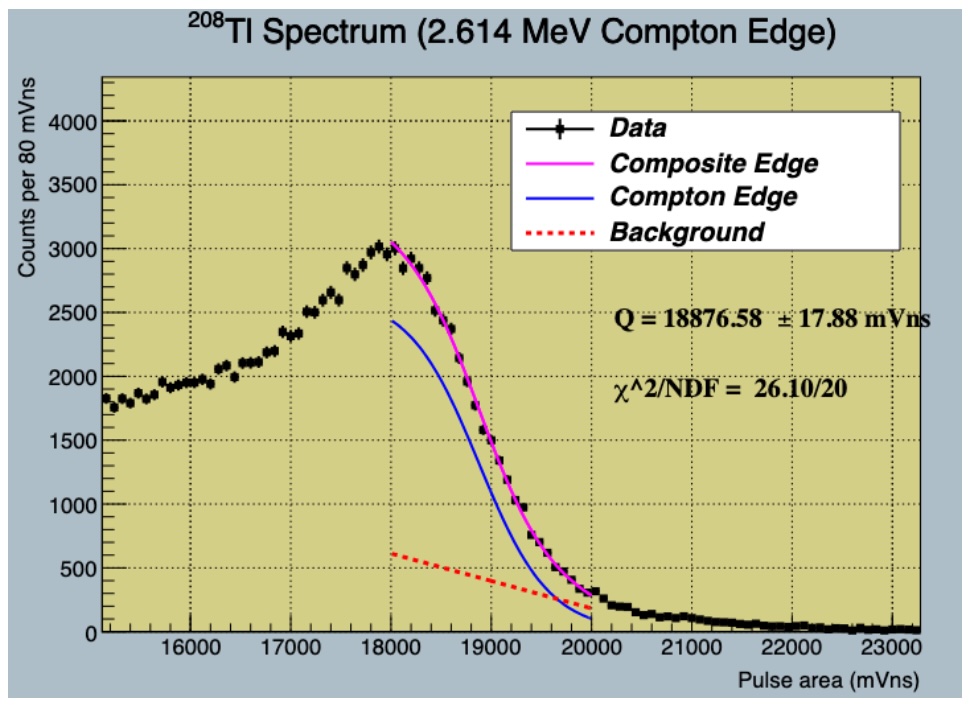}
\includegraphics[width=.5\textwidth]{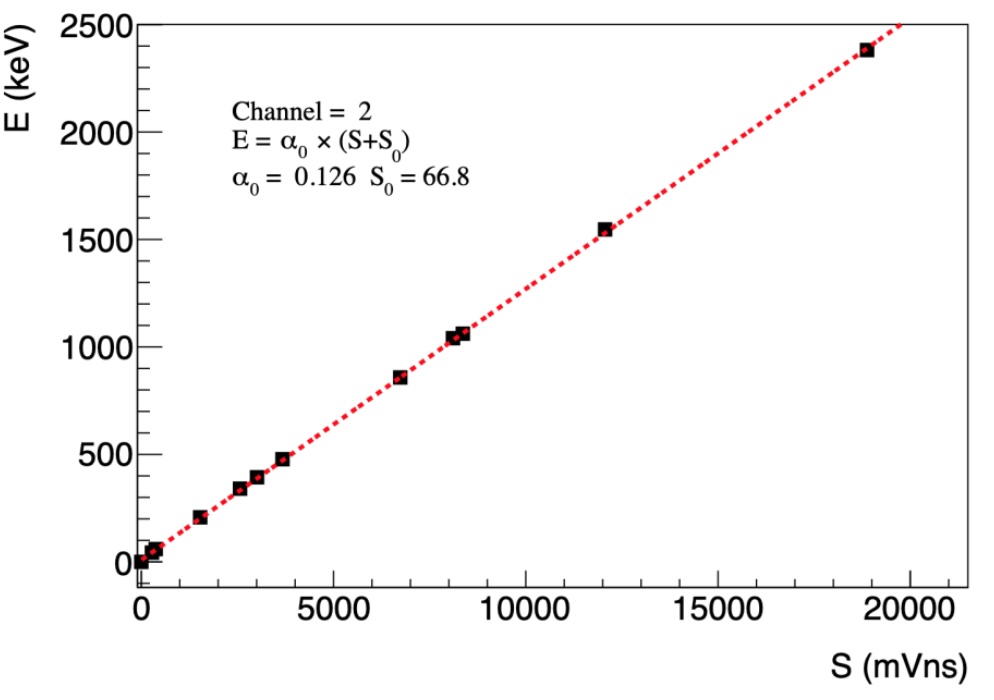}
\caption{(Left) $^{208}$Tl Compton edge spectrum obtained with
a single ALMOND module. (Right) An example pulse area (S) vs. energy (E)
calibration curve.}
\label{fig:Gamma}
\end{figure}

ALMOND calibration program
at Karlsruhe Institute of Technology (KIT)
was broken down into three categories,
namely gamma calibration, proton recoil calibration, and neutron
capture calibration. First, each of 36 detector segments
was individually calibrated with various gamma sources.
These sources include $^{241}$Am (59.5 keV), $^{133}$Ba (0.356 MeV),
$^{22}$Na (0.511 MeV and 1.275 MeV), $^{207}$Bi (0.570 MeV, 1.064 MeV
and 1.770 MeV), $^{137}$Cs (0.662 MeV), $^{60}$Co (1.17 MeV and 1.33 MeV)
and $^{232}$Th (2.614 MeV). Due to the low density of PS, none of the gammas except for $^{241}$Am were fully converted but resulted in
Compton edge signatures.
Figure~\ref{fig:Gamma} (left) displays an example Compton spectrum.
Each Compton spectrum was empirically fitted with a Fermi function~\cite{Adhikari_2018} combined with a linear background. Calibration
points extracted from the Fermi function were put together
to establish gamma energy calibration curves, which exhibit
a linear behavior as shown in figure~\ref{fig:Gamma} (right).

\begin{figure}[htbp]
\centering
\includegraphics[width=.45\textwidth]{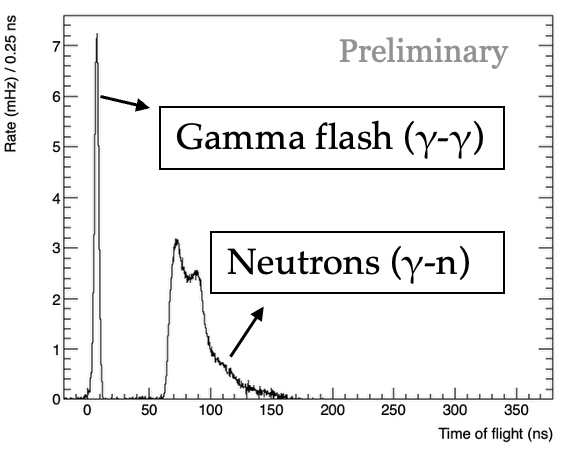}
\includegraphics[width=.39\textwidth]{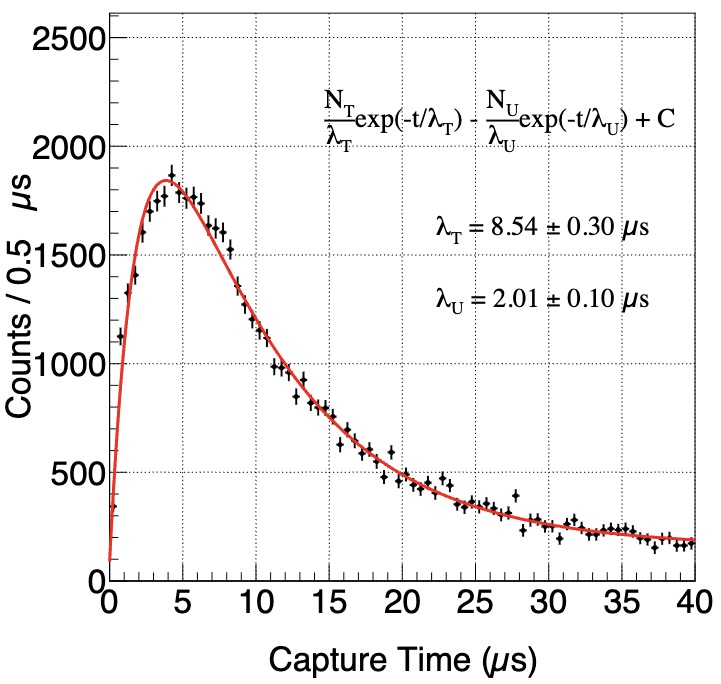}
\caption{(Left) ToF spectrum measured with the tagged
neutron source and an individual ALMOND module. The first peak represents
the $\gamma$-$\gamma$ coincidence between the BGO detector and the PS
unit, respectively, whereas the later distribution
shows the neutrons, denoted by $\gamma$-n coincidence.
(Right) Time profile of the tagged AmBe neutrons captured by ALMOND.
The capture time distribution was empirically fitted with a
double exponential function~\cite{Littlejohn:2012qpa}.
$\lambda_U$ depicts the rising edge owing to the thermalization of
fast neutrons and $\lambda_T$ refers to the falling edge
due to the capture of thermalized neutrons.}
\label{fig:Neutron}
\end{figure}

For proton recoil calibration, we employed a method known as
Time-of-Flight (ToF). Using a gamma emitting AmBe (Americium-Beryllium)
neutron source
placed right next to a gamma detector (BGO), a single ALMOND module
was located at about 2\,m of distance. A $\gamma$-pulse in the BGO detector
tagged the late arrival of the correlated AmBe neutron to the ALMOND module. It can be inferred from figure~\ref{fig:Neutron} (left) that neutrons
were trivially identified in the ToF spectrum. Given the fixed distance,
the neutron energies were derived from the neutron ToFs. The tagged
AmBe source characterization was accomplished in this step.
Next, the neutron energy spectrum was divided into multiple energy bins.
For each bin, the end point of the recoil spectrum corresponds
to the case where the neutron loses all of its energy in a single scatter.
Then, the Birks parameter was empirically obtained, which is in agreement
with another similar ToF measurement in the literature~\cite{Langford:2015fea}.

The same tagged neutron source setup was used to calibrate the neutron
capture in the complete detector. The source distance was set to 50\,cm,
and the BGO detector was again placed very close to the source. In
this scheme, the proton recoils were selected by the
coincident BGO pulse, followed by the neutron capture pulse in ALMOND
within 40\,$\mu$s. This campaign enables us to construct the spectral unfolding
of proton recoils and also 
determine the neutron capture time profile,
as shown in figure~\ref{fig:Neutron} (right).
Furthermore, a series of calibration
measurements were carried out at the Frascati Neutron Generator (FNG)
facility~\cite{Frascati2018} using a monoenergetic DD neutron generator
and a well-characterized AmB (Americium-Boron) neutron source.

\section{Commissioning at the LNGS underground laboratory}

\begin{figure}[htbp]
\centering
\includegraphics[width=.4\textwidth]{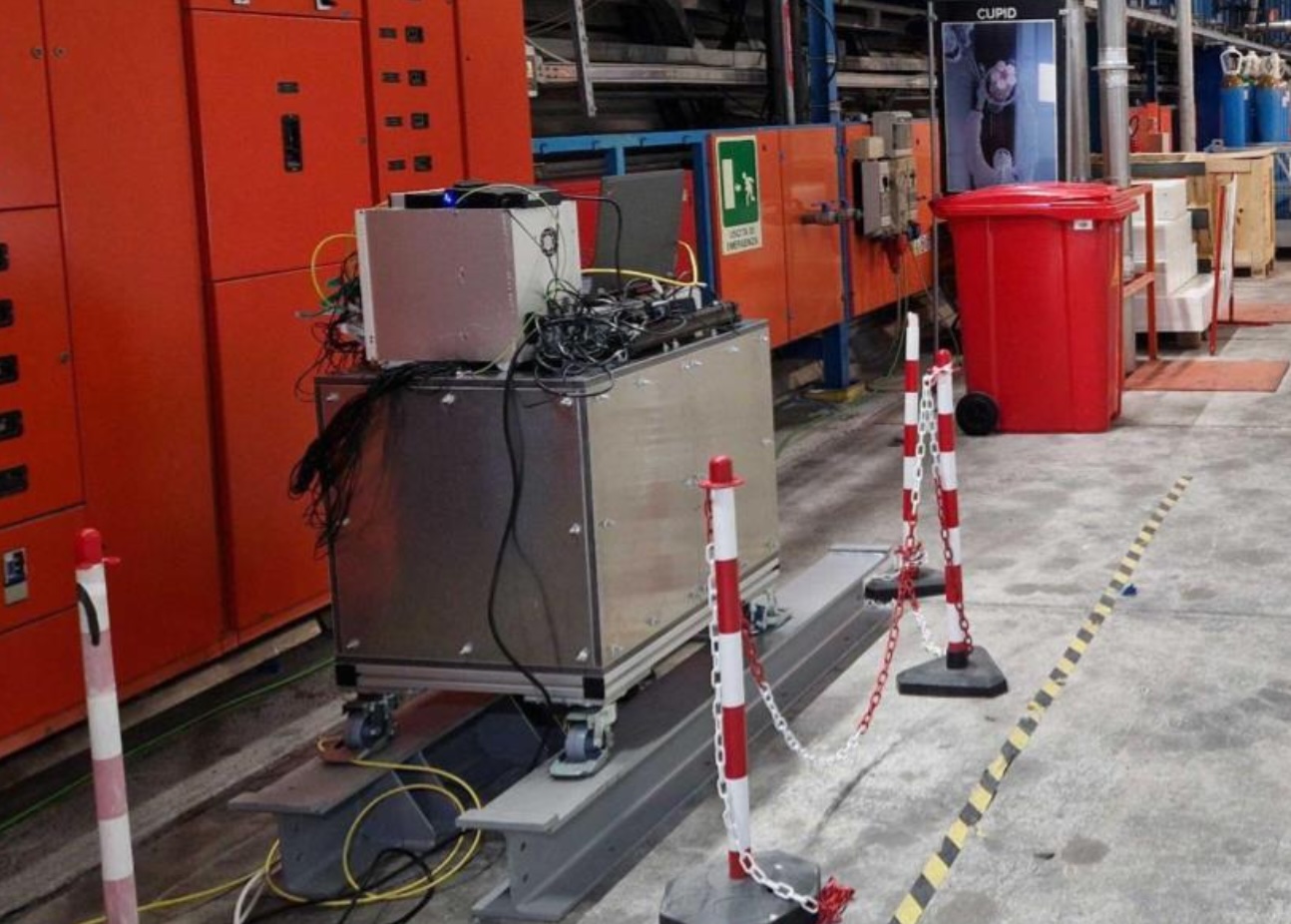}
\caption{ALMOND positioned in Hall A of the LNGS underground laboratory during a first long-term data taking.}
\label{fig:Commission}
\end{figure}

ALMOND was commissioned in Hall~A of the LNGS underground laboratory,
as shown in figure~\ref{fig:Commission}. Initially, an ambient
gamma measurement was conducted to estimate the background rate.
This was calculated by multiplying the fake capture rate ($>$3 MeV)
in this data by the fake proton recoil rate ($>$20 keV$_\mathrm{ee}$)
and the pre-trigger time window (40\,$\mu$s).
We concluded that the background event rate
would be subdominant in comparison to the expected neutron rate.

\begin{figure}[htbp]
\centering
\includegraphics[width=.8\textwidth]{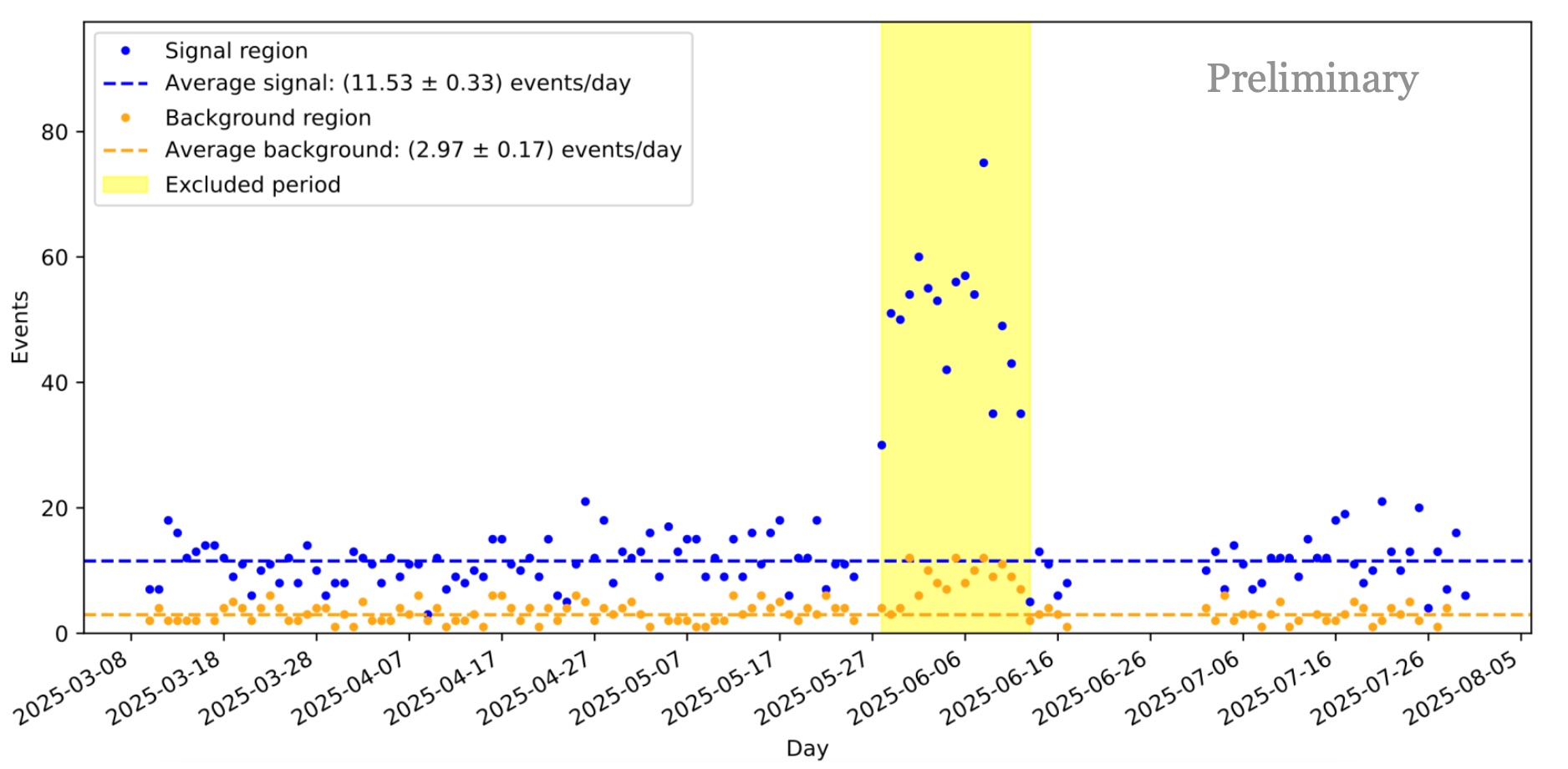}
\caption{Result in units of count rate per day of the ambient neutron measurement in Hall A}
\label{fig:result}
\end{figure}

The ambient neutron measurement in Hall~A lasted more than 3 months.
Figure~\ref{fig:result} shows the event counts day-by-day.
The blue dots display the signal region events, where the main
trigger pulse was preceded by another pulse, as anticipated from true
capture events. The orange dots, on the other hand, represent the accidental background
region events, where the trigger pulse was matched with a pulse
in the post-trigger time window. The signal region counts were consistently
larger than the background region counts, indicating neutron detection.
The preliminary analysis suggests that the average rate in the signal
and background regions were $11.53\pm0.33$ and $2.97\pm0.17$ events/day,
respectively, resulting in a detected neutron rate of
$8.6\pm0.4$ events/day. The excess rate in the yellow band
was due to a neutron calibration campaign in a nearby experiment.
This period was excluded from the analysis. The gap following the yellow
band indicates a period without measurement.

\section{Conclusion and outlook}
ALMOND was designed to measure the low-flux ambient neutron fields 
at the LNGS underground laboratory.
The initial measurements in Hall~A show that the ALMOND design was a success.
The ambient neutron measurement in Hall~A was completed
and the neutron survey in Hall~C is currently ongoing. We
are conducting detailed studies to benchmark the
neutron efficiency using the calibration datasets taken at the KIT
and FNG facilities. The short-term plan is to estimate the
neutron fluxes in Hall~A and Hall~C.

\acknowledgments{
We acknowledge the financial support from the German Federal Ministry of Research, Technology and Space (BMFTR) under the grant number 05A21VK1 and the Italian National Institute for Nuclear Physics (INFN). We thank our colleagues at ENEA Frascati for providing access and technical assistance during the ALMOND calibration at the Frascati Neutron Generator.
}

\bibliographystyle{JHEP}
\bibliography{references}

\end{document}